\documentclass[12pt]{article}

\usepackage{amsmath,amssymb,bm,graphicx,natbib,fullpage}

\begin{document}

\title{Discussion of ``Sequential Quasi-Monte Carlo'' by Mathieu Gerber and Nicolas Chopin}

\author{\large Dr. Chris. J. Oates\footnote{Email: c.oates@warwick.ac.uk}, \; Dr. Daniel Simpson\footnote{Email: d.p.simpson@warwick.ac.uk} \; and Prof. Mark Girolami\footnote{Email: m.girolami@warwick.ac.uk} \\ Department of Statistics, Zeeman Building, University of Warwick,\\ Gibbet Hill Road, Coventry, CV4 7AL, UK}

\date{}

\maketitle

This paper is timely for highlighting the benefits of Quasi-Monte Carlo (QMC) in contemporary computational statistical methodology.
Below we address the question of whether there is scope to further reduce the error of QMC estimators.
The analysis of QMC used by Gerber and Chopin is rooted in the Koksma-Hlawka inequality
\begin{eqnarray*}
\left| \frac{1}{N} \sum_{n=1}^N \varphi(\bm{u}^n) - \int_{[0,1]^d} \varphi(\bm{u}) d\bm{u} \right| \leq V(\varphi) D^*(\bm{u}^{1:N})
\end{eqnarray*}
where $\varphi : [0,1]^d \rightarrow \mathbb{R}$ is a test function of interest,  $\bm{u}^{1:N}$ is a point set (or sequence), $V(\varphi)$ is the (Hardy-Krause) total variation and $D^*(\bm{u}^{1:N})$ is the (star) discrepancy term that is the target of the QMC innovation.
Our discussion explores the potential to simultaneously tackle the rate constant $V(\varphi)$ in conjunction with the use of QMC methods to tackle $D^*(\bm{u}^{1:N})$.
This direction has received considerably less attention due to typical analytic intractability of the rate constant. 
\cite{Hickernell} showed that classical control variate strategies from Monte Carlo (MC) are typically not well-suited to QMC, since the total variation is only weakly related to the MC variance that is the target of classical variance reduction techniques.
Below we hint toward a general strategy to reduce QMC error that targets the rate constant directly.

Following recent work on ``control functionals'' by \cite{Oates}, we consider evaluation of $\varphi$ on two sets $\bm{u}^{1:N}$ and $\bm{v}^{1:N}$ at a computational cost (asymptotically) equivalent to evaluating $\varphi$ on one such set.
The first set $\bm{u}^{1:N}$ is used to compute an arithmetic mean
\begin{eqnarray*}
I_{\text{CF}} = \frac{1}{N} \sum_{n=1}^N \hat{\varphi}_N(\bm{u}^n), \label{est}
\end{eqnarray*}
based on a surrogate function $\hat{\varphi}_N : [0,1]^d \rightarrow \mathbb{R}$. 
This surrogate function is itself estimated from the second set $\bm{v}^{1:N}$, in a preliminary step.
In situations where $\hat{\varphi}_N$ can be made to satisfy (i) $\int\hat{\varphi}_N(\bm{u})d\bm{u} = \int\varphi(\bm{u})d\bm{u}$ for all $N \in \mathbb{N}$ and (ii) $V(\hat{\varphi}_N) \rightarrow 0$ as $N \rightarrow \infty$, then the control functional estimator $I_{\text{CF}}$ is unbiased (in an appropriate sense) and has asymptotically zero error relative to the standard QMC estimator.
\cite{Oates2} provides an explicit implementation of this strategy in the more general reproducing kernel Hilbert space formulation of QMC methodology \citep{Dick}.

\begin{figure}
\centering
\includegraphics[width = 0.7\textwidth,clip,trim = 3.5cm 8.5cm 3.5cm 8.5cm]{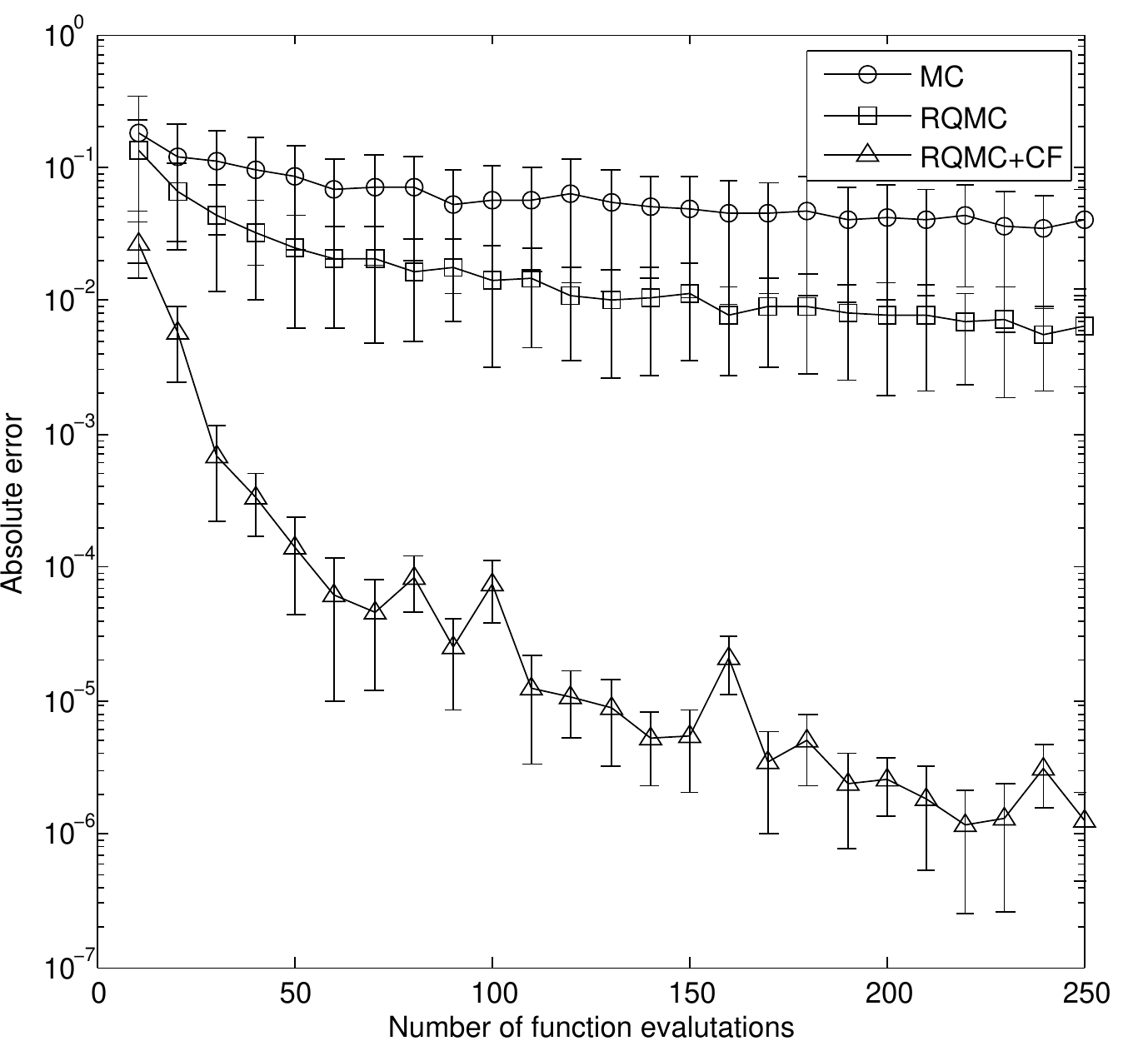}
\caption{Convergence of control functional (CF) Quasi-Monte Carlo (QMC) estimators with respect to more standard alternatives.
This example considers the (one-dimensional) test function $\varphi(x) = \sin(2\pi x) + 4x$.
The usual Monte Carlo (MC) rate is $O(N^{-1/2})$ and the usual QMC rate is $O(N^{-1+\epsilon})$ (for any $\epsilon>0$).
The QMC+CF approach has a rate that is $O(N^{-2+\epsilon})$.
[Here we present the Randomised QMC (RQMC) case: 
For $\bm{u}^{1:N}$ we used a scrambled Halton sequence of length $N$ with a uniform random shift modulo one and for $\bm{v}^{1:N}$ we used a uniform grid on the unit interval.
Error bars denote $\pm 1$ standard deviation.
For a fair comparison, the same number of function evaluations for $\varphi$ was used in evaluating each estimator.
It is striking how much the estimation error can be reduced by using control functionals.]}
\label{example}
\end{figure}

As a simple example, we note that for differentiable $\varphi$ with sufficiently regular partial derivatives, a basic implementation produces a total variation $V(\varphi_N)$ that vanishes at a rate $O(N^{-1/d})$. 
Thus control functional QMC estimators are asymptotically superior to standard (R)QMC estimators under appropriate regularity conditions.
Preliminary empirical results strongly support our theoretical analysis; an example is given in Fig. \ref{example}.

Given the gains in accuracy that are provided by QMC, it is surely a priority to establish complementary methodology that targets the rate constant governing the practical performance of these algorithms.
Control functionals provide one (explicit) route to achieve this goal.
The combination of control functionals with the Sequential QMC approach of Gerber and Chopin should provide a highly effective approach to estimation.

\end{document}